\documentstyle[epsfig,latexsym,amssymb,aps,floats,eqsecnum,preprint,
amsfonts]{revtex}
\def\laq{\raise 0.4ex\hbox{$<$}\kern -0.8em\lower 0.62 ex\hbox{$\sim$}}
\def\gaq{\raise 0.4ex\hbox{$>$}\kern -0.7em\lower 0.62 ex\hbox{$\sim$}}

\def\vk{\vec{k}}
\def\vp{\vec{p}}
\def\vx{\vec{x}}
\def\vy{\vec{y}}

\begin{document}
\begin{titlepage}
\begin{flushright}
CERN-TH/2003-185
\end{flushright}
\vspace*{1cm}

\begin{center}
{\large{\bf Assigning  Quantum-Mechanical Initial Conditions to Cosmological Perturbations}}
\vskip 1 cm 
{\sl Massimo Giovannini\footnote{Electronic address: massimo.giovannini@cern.ch}}
\vskip 0.5 cm 
{\sl  Theoretical Physics Division, CERN, CH-1211 Geneva 23, Switzerland}
\vspace{1cm}
\noindent
\begin{abstract}
Quantum-mechanical initial conditions for the fluctuations 
of the geometry can be assigned in excess of a 
given physical wavelength. The two-point 
functions of the scalar and tensor modes of the geometry 
will then inherit corrections depending on which Hamiltonian 
is minimized at the initial stage of the evolution. 
The energy density of the background geometry is compared 
with the energy-momentum pseudo-tensor of the fluctuations
averaged over the initial states, minimizing each different 
Hamiltonian.
The minimization of adiabatic Hamiltonians leads to initial 
states whose back-reaction on the geometry is negligible. 
The minimization of non-adiabatic Hamiltonians, 
ultimately responsible for  large corrections in the two-point 
functions, is associated with initial states whose 
energetic content is of the same order as the energy density 
of the background.
\end{abstract} 
\vskip0.5pc
\end{center}
\end{titlepage}
\newpage
\noindent

\renewcommand{\theequation}{1.\arabic{equation}}
\setcounter{equation}{0}
\section{Introduction} 
In cosmology, classical and quantum fluctuations share 
some features, which can  hide radical differences.
For instance, in the linearized approximation, classical 
and quantum fluctuations obey the  
same evolution equations, but  while classical fluctuations are given 
once forever (on a given space-like hypersurface) quantum fluctuations 
keep on reappearing all the time during the inflationary phase.

Inflation has to last approximately $60$-efolds. One reason to demand  
such a minimal duration is that, today, the total curvature of the Universe 
receives a leading contribution from the extrinsic curvature and a subleading 
contribution from the intrinsic (spatial) curvature. The ratio between the 
intrinsic and extrinsic curvature goes as $1/\dot{a}^2$ 
(where $a(t)$ is the scale 
factor of the Friedmann--Robertson--Walker Universe and the dot denotes derivation 
with respect to the cosmic time coordinate). During an epoch of 
decelerated expansion (i.e. $\ddot{a}<0$, $\dot{a} >0$) such as 
the ordinary radiation  and matter-dominated phases, $1/\dot{a}^2$ can become 
very large. The r\^ole of inflation is, in this context, to 
make $1/\dot{a}^2$ very minute at the end of inflation, so that it can easily be 
of order 1 today. The minimal duration of inflation required 
in order to achieve this 
goal is about $60$-efolds.

If the duration of inflation is minimal (or close to minimal) 
classical fluctuations, which were super-horizon sized at the onset
 of inflation will 
be affected neither by the inflationary phase nor by the subsequent 
post-inflationary epoch and can have computable large scale effects 
\cite{GZ,GZ2}. 
If the fluctuations are  {\em classical},
there are, virtually no ambiguities in  normalizing them: it is sufficient 
to assign the values of the various inhomogeneities over a 
typical scale and at a given time. For instance, one can imagine 
that at the onset of inflation the tensor modes of the geometry 
had some classical initial conditions; this observation leads to predictable
consequences {\em provided} the duration of inflation is minimal \cite{GGV1}.

When the duration of inflation is 
much longer than $60$-efolds, the large scale 
fluctuations are probably all of quantum-mechanical nature, 
at least in the case of inflationary models driven by a single inflaton field.
Quantum-mechanical fluctuations result 
from the zero-point energy of the metric inhomogeneities 
present during the inflationary epoch. 
The  predictions of inflationary cosmology are partially imprinted 
in the correlation functions of the scalar and tensor modes of the  
geometry. For a reliable calculation of these correlators, it is 
mandatory to correctly normalize the inhomogeneity 
of the geometry to their quantum-mechanical value.

The main theme of the present investigation  will be to present
various  ways of assigning quantum-mechanical initial 
conditions in the treatment of cosmological perturbations. 
The leading term of the scalar and tensor power spectra will not be 
affected by the different prescriptions. However, there 
will be computable corrections, which change according
to the choice of normalization assignment.
The second step of the present paper will be to select one of the different 
 prescriptions, according to the requirement that 
the initial state of the evolution of the 
fluctuations will not carry too much energy density if compared 
with the background geometry. 

A naive (but correct) answer to the problem of normalizing 
quantum fluctuations is  
to demand that the initial state for the evolution 
of the various modes of the geometry minimizes the corresponding (quantum) 
Hamiltonian at the onset of the time evolution.
In spite of the correctness of the previous statement, ambiguities 
are hidden in so far as the Hamiltonian is time-dependent.

One way of assigning quantum-mechanical initial 
conditions consists in normalizing  the  mode functions to their 
``vacuum'' value for \footnote{In this paper
$\eta$ denotes the conformal time cooordinate, simply related to the cosmic 
time coordinate by the standard differential relation $a(\eta) d\eta = d t$. } $\eta \to -\infty$. 
The Hamiltonians 
of the scalar and tensor modes of the geometry will then 
be minimized for  $\eta \to -\infty$, which  is a physical limit, not a mathematical one. 
In fact, inflation cannot last indefinitely in the past. Thus,  
metric  fluctuations are normalized to their quantum mechanical amplitude 
at a time very close to the onset of inflation.

The standard way of normalizing the fluctuations of the geometry was recently 
scrutinized in different contexts (see \cite{GV} and  \cite{br1} for  recent reviews covering 
also this subject). The approach of these investigations is from different perspectives.
In  \cite{star,dani,max1} (see also \cite{kal,picon,chung}), 
the robustness of inflationary predictions is discussed 
from a conservative point of view. In \cite{br2,east,nim,sloth}  modifications 
of the dispersion relations (arising from different contexts) are invoked as a 
possible source of deviation from the standard lore.

In order to explain in simple terms why the standard prescription {\em might} 
be questioned, let us consider the situation where inflation lasts more than the 
(minimal) $60$-efolds.
Consider also, for concreteness, the case of the tensor modes of the 
geometry in  de Sitter space described by a 
scale factor $a(\eta) = (-\eta_1/\eta)$, for $\eta \leq -\eta_1$. In this case 
the evolution equation for each tensor polarization is particularly simple and it is given by:
\begin{equation}
h_{k}'' -\frac{2}{\eta}h_{k}' + k^2 h_{k} =0.
\label{a}
\end{equation}
This equation of motion can be obtained from different Hamiltonians. For instance, following 
\cite{max1} and defining $\mu = a h$ we will have
\begin{eqnarray}
H(\eta) &=& \frac{1}{2} \int d^{3}x 
\biggl[ \pi^2  - \frac{2}{\eta} \mu \pi + (\partial_{i} \mu)^2 \biggr],
~~~~~~~ \pi = \mu' + \frac{1}{\eta} \mu,
\label{c}\\
\tilde{H}(\eta) &=& \frac{1}{2} \int d^{3}x 
\biggl[ \tilde{\pi}^2  - \frac{2}{\eta^2} \mu^2 
+ (\partial_{i} \mu)^2 \biggr],
~~~~~~~~~\tilde{\pi} = \mu'. 
\label{d}
\end{eqnarray}
Using the explicit expression for the canonical momenta, the Hamilton equations derived from either 
(\ref{c}) or (\ref{d}) will always lead, when combined, to (\ref{a}). 
It is expected that the two Hamiltonians will lead  to exactly the same dynamical evolution,
since (\ref{c}) and (\ref{d})  are related by a canonical transformation. Furthermore, 
in the  limit
$\eta \to -\infty$, Eqs. (\ref{c}) and (\ref{d}) coincide, since  $\tilde{\pi} \sim \pi$. 
Consequently,  if, as in the standard treatment,  quantum-mechanical initial 
conditions are assigned for 
$\eta \to -\infty$,  the state minimizing $H$ will also minimize $\tilde{H}$. 
On the contrary, when initial conditions are imposed at   
 a {\em finite} (but large) conformal time $\eta_{0}$,  
the states minimizing $H$ and $\tilde{H}$ will differ. 
This 
is, ultimately, the reason why \cite{max1}  the various
authors in \cite{star,dani} get different corrections 
in the (late-time) two-point 
function of $h(\vx,\eta)$. Different Hamiltonians 
(not necessarily coinciding with (\ref{c}) or (\ref{d}))
are minimized at the initial time of the evolution: while 
the dynamical evolution is the same for both Hamiltonians, 
the quantum-mechanical states 
minimizing one or the other are different.

Concerning the initial time of the evolution
of the fluctuations there are two possibilities: it could be independent 
of the comoving scale or it could be different  
depending on the comoving scale. 
In order to illustrate the first possibility, recall 
 that, in the conformal time parametrization 
the scale factor $a(\eta) \to 0$ for $\eta \to -\infty$. 
Thus, the  physical wavelength 
\begin{equation}
\lambda_{\rm ph}(\eta) = \lambda_0 a(\eta)
\end{equation}
of the fluctuations  will go to zero for $\eta \to -\infty$.
The typical amplitude of the tensor fluctuations is given by the power spectrum, 
i.e. the Fourier transform 
of the two-point function which is, up to numerical factors 
$\delta_{h} \sim k^{3/2} |h_{k}|$.
The fluctuations described by Eq. (\ref{a}) should be normalized to quantum-mechanical fluctuations 
{\em before} they leave the horizon, i.e. in the regime $ k \eta \gg 1$. In this regime, 
$h_{k} \sim 1/a$ and 
the fluctuation is said to be adiabatically suppressed.
It is clear that for a  quantum-mechanical fluctuation, i.e.  $|h_{k}|\sim \ell_{\rm P}/\sqrt{2 k}$ 
we will have  $\delta_{h} 
\sim  \omega(\eta)/M_{\rm P}$, where $\omega(\eta) = \lambda(\eta)^{-1}= k/a(\eta) $ is the 
physical frequency. It is also  clear that when $ \omega(\eta) \sim M_{\rm P}$, 
$\delta_{h} \sim {\cal O}(1)$.

There might 
be nothing wrong with the fact that $\lambda_{\rm ph}$ goes to zero; however one can also 
imagine that the physical description contains a fundamental length scale $\Lambda^{-1}$.
In this case the time at whivh the normalization is assigned changes depends on the 
physical scale.
Suppose then that a quantum-mechanical normalization is assigned 
to the fluctuations, at a given conformal time $\eta_0$ in  such a way that the physical 
wavelength
\footnote{We wiill denote with $\lambda_{\rm ph}(\eta)$ the physical 
wavelength and  with $\omega(\eta)= k/a(\eta)$ 
the physical fcrequency} is defined as , 
\begin{equation}
\lambda_{\rm ph}(\eta_0) = \Lambda^{-1},
\label{cond1}
\end{equation}
 where $\Lambda^{-1}$ is a typical
length scale which is of the order of the Planck scale \cite{star,dani,GV}. 
The condition (\ref{cond1}) defines a  New Physics Hypersurface (NPH) 
\cite{max1}, in the sense that, unlike in the standard case, different physical
frequencies become of order $\Lambda$ at different conformal times.
 
In the first part of this investigation the corrections to the power 
spectrum of scalar and tensor fluctuations of the geometry will be computed. 
It will be shown 
 that the minimization of different Hamiltonians, 
characterized by a different degree of adiabaticity, 
lead to different corrections to the power spectrum 
of curvature fluctuations. The same ambiguities 
arising in the case of the tensor modes of the geometry \cite{max1} 
are also present in the case 
of the curvature and metric fluctuations.
 
In general one cannot assign a localized energy density 
to the gravitational field, and this is one of the problems 
in the analysis of the back-reaction of gravitational fluctuations.
One possible approach is the one of \cite{br3} (see, 
for a different perspective, also \cite{gr3}). The energy density 
of tensor fluctuations of the geometry will be estimated for different 
initial states (minimizing different Hamiltonians) and it will be shown
that this analysis  pins down a specific class of Hamiltonians.

The present paper is organized as follows. In Section II  
the main equations for the scalar modes of the geometry will be briefly reviewed. In
Section III it will be shown that 
minimizing different Hamiltonians will lead to different corrections in the 
scalar power spectra.  Then, in Section IV
 the issue of the selection of the Hamiltonian will be addressed. 
The energy density of the tensor modes off the geometry (derived from the  
pseudo-tensor of gravitational waves) will be averaged over the quantum states, minimizing different 
Hamiltonians. It will be shown that  different initial states  can be distinguished by requiring 
that their energetic content is always sub-leading with respect to the energy density of the 
background geometry.
Finally in Section V some concluding remarks will be proposed.

\renewcommand{\theequation}{2.\arabic{equation}}
\setcounter{equation}{0}
\section{Curvature and metric fluctuations} 
Consider an accelerated phase of expansion driven by a single inflaton field $\varphi$
in a spatially flat Friedmann--Robertson--Walker 
metric whose line element can be written, in the conformal time 
parametrization, as 
\begin{equation}
ds^2 = a^2(\eta)[d\eta^2 - d\vx^2].
\end{equation}
The evolution equations for the background will then be
\footnote{If not stated otherwise, 
units $M_{P} =  1/\sqrt{8 \pi G}$ will be used.} 
\begin{eqnarray}
&&{\cal H}^2 M_{\rm P}^2 = \frac{1}{3} \biggl( \frac{{\varphi'}^2}{2 } + V a^2 \biggr),
\label{b1}\\
&&2 ( {\cal H}^2 - {\cal H}') M_{\rm P}^2  = \varphi'^2 ,
\label{b2}\\
&& \varphi'' + 2 {\cal H} \varphi' + \frac{\partial V}{\partial \varphi} a^2 =0,
\label{b3}
\end{eqnarray}
where the prime denotes a derivation with respect to the conformal time coordinate and ${\cal H} = a'/a$.

The two Bardeen \cite{bard} potentials ($ \Psi$ and $\Phi$) and the 
gauge-invariant scalar field fluctuation $\chi$ define the coupled system 
of scalar fluctuations of the geometry (see, for instance, \cite{max2}):
\begin{eqnarray}
&&
\nabla^2 \Psi - 3 {\cal H} ( {\cal H} \Phi + \Psi') = \frac{a^2}{2 M_{\rm P}^2} \delta\rho_{\varphi},
\label{00}\\
&&  {\cal H} \Phi + \Psi'  = \frac{1}{2 M_{\rm P}^2} \varphi' \chi,
\label{0i}\\
&& \Psi'' + {\cal H} (\Phi' + 2  \Psi') 
 + ({\cal H}^2 + 2 {\cal H}') \Phi  = \frac{ a^2}{2 M_{\rm P}^2} \delta p_{\varphi}
\label{ij} 
\end{eqnarray}
where Eqs. (\ref{00})--(\ref{ij}) are, respectively, the perturbed $(00)$, $(0i)$ and $(ij)$ components 
of Einstein equations and  
\begin{eqnarray}
\delta\rho_{\varphi} &=& \frac{1}{a^2} 
\biggl[ - \varphi'^2 \Phi + \varphi' \chi' + \frac{\partial V}{\partial \varphi} \chi\biggr],
\nonumber\\
\delta p_{\varphi} &=& \frac{1}{a^2} \biggl[ - \varphi'^2 \Phi + \varphi' \chi'
 - \frac{\partial V}{\partial \varphi}\chi \biggr],
\end{eqnarray}
are the gauge-invariant energy and pressure density fluctuations.
In the absence of any shear in the perturbed energy-momentum tensor) 
the $(i \neq j)$ component of the perturbed Einstein equations 
leads to $\Phi = \Psi$.

Following \cite{bard} and \cite{lyth}, it is convenient to introduce
 the fluctuations of the spatial curvature on comoving 
spatial hypersurfaces
\begin{equation}
{\cal R} = - \Psi - {\cal H} \frac{\chi}{\varphi'}= 
- \Psi - \frac{{\cal H}( {\cal H} \Phi + \Psi')}{{\cal H}^2 - {\cal H}'},
\label{defR}
\end{equation}
where the equality follows from the use of Eq. (\ref{0i}) and of the background equations.
The definition of (\ref{defR}) and a linear combination of Eqs. (\ref{00}) and (\ref{ij}) 
leads to the following simple equation
\begin{equation}
{\cal R}' = - \frac{4 {\cal H}}{{\varphi'}^2} \nabla^2 \Psi,
\label{eqR}
\end{equation}
which implies the constancy of ${\cal R}$ for modes $k\eta \ll 1$ \cite{bard,lyth}.
The power spectrum 
of the scalar modes amplified during the inflationary phase is customarily 
expressed in terms of ${\cal R}$, which is conserved on super-horizon scales.
Taking the time 
derivative of Eq. (\ref{eqR}) and using, repeatedly, Eq. (\ref{defR}) and 
Eqs. (\ref{00})--(\ref{ij}), the following second-order equation can be obtained:
\begin{equation}
{\cal R}'' + 2 \frac{z'}{z} {\cal R}' - \nabla^2 {\cal R} =0,
\label{seceq}
\end{equation}
where 
\begin{equation}
z = \frac{a \varphi'}{\cal H}.
\end{equation}
Going to Fourier space, Eq. (\ref{seceq}) has a simple solution 
for modes outside the horizon, i.e.
\begin{equation}
{\cal R}_{k} = A_{k} + B_{k} \int^{\eta} \frac{d\eta'}{z^2(\eta')},
\end{equation}
namely, for the case of single field inflationary backgrounds with polynomial
or exponential potential, a constant and a decaying solution.

The curvature perturbations on comoving spatial hypersurfaces can also 
be simply related to the curvature perturbations 
on the constant density hypersurfaces, denoted by $\zeta$
\begin{equation}
\zeta = - \Psi - {\cal H} \frac{\delta\rho_{\varphi}}{\rho_{\varphi}'} \equiv 
- \Psi + \frac{a^2 \delta\rho_{\varphi}}{3 {\varphi'}^2}.
\label{defzeta}
\end{equation}
It is clear that, taking  the difference in the definitions 
(\ref{defR}) and (\ref{defzeta}), and using Eq. (\ref{00}):
\begin{equation}
\zeta - {\cal R} \equiv {\cal H}\frac{\chi}{{\varphi}'} + \frac{ a^2 \delta \rho_{\varphi}}{3 {\varphi'}^2} = 
\frac{ 2 M_{\rm P}^2}{3} \frac{\nabla^2 \Psi}{{\varphi '}^2} ,
\end{equation}
${\cal R}$ and $\zeta$ differ by Laplacians of the Bardeen potential.

For the purposes of the present investigation, it is desirable to treat 
the evolution of the scalar fluctuations of the geometry in terms of 
a suitable variational principle. On this basis consistent Hamiltonians 
for the evolution of the fluctuations can be defined.

Instead of perturbing the Einstein equations to first order,
the Einstein-Hilbert and  scalar field actions should be perturbed 
to second order. The result of this procedure 
is usefully expressed in terms 
of the gauge-invariant curvature fluctuation:
\begin{equation}
 S^{(1)} = \frac{1}{2} \int d^4 x\,\, z^2 \biggl[{ {\cal R}'}^2 - (\partial_{i} {\cal R})^2\biggr].
\label{Raction}
\end{equation}
Defining now the canonical momentum $\pi_{\cal R} = z^2 {\cal R}'$ the Hamiltonian related 
to the action (\ref{Raction}) becomes
\begin{equation}
H^{(1)}(\eta) = \frac{1}{2}\int d^{3}x\,\, 
\biggl[ \frac{\pi_{{\cal R}}^2}{z^2} + z^2  (\partial_{i} {\cal R})^2 \biggr],
\label{hamscal1}
\end{equation}
and the Hamilton equations 
\begin{eqnarray}
&& \pi_{\cal R}' =  z^2 \nabla^2 {\cal R},
\label{hameq1}\\
&& {\cal R}' = \frac{\pi_{{\cal R}}}{z^2}.
\label{hameq2}
\end{eqnarray}
Combining these equations in a single second-order equation 
 Eq. (\ref{seceq}) is again obtained.

The physical interpretation of ${\cal R}$ has been already introduced
in terms of the curvature fluctuations on comoving spatial hypersurfaces. The 
 canonically conjugate momentum, $\pi_{{\cal R}}$ 
is related to the  
density contrast on comoving hypersurfaces, namely, in the case of a single scalar field 
source\cite{bard},
\begin{equation}
\epsilon_{\rm m} = \frac{ \delta \rho_{\varphi} + 3 {\cal H} (\rho_{\varphi} + p_{\varphi}) V}{\rho_{\varphi}} = 
\frac{a^2\delta\rho_{\varphi} + 3 {\cal H}\varphi' \chi}{a^2 \rho_{\varphi}},
\label{epsm}
\end{equation}
where the second equality can be obtained using that $\rho_{\varphi} + p_{\varphi} = {\varphi'}^2/a^2 $ and that the 
effective ``velocity'' field in the case of a scalar fiedl is $ V = \chi/\varphi'$.
Making now use of Eq. (\ref{0i}) into Eq. (\ref{00}), Eq. 
(\ref{epsm}) can be expressed as  
\begin{equation}
\epsilon_{\rm m} = \frac{ 2 M_{\rm P}^2 \nabla^2 \Psi}{a^2 \rho_{\varphi}} \equiv
 \frac{2}{3}\frac{\nabla^2\Psi}{{\cal H}^2},
\label{epm2}
\end{equation}
where the last equality follows from  Eq. (\ref{b1}).

From Eq. (\ref{epm2}), it also follows that 
\begin{equation}
\pi_{{\cal R}} = z^2 {\cal R}' \equiv - 6 a^2 {\cal H} \epsilon_{\rm m},
\label{piR}
\end{equation}
where Eq. (\ref{epm2}) has been used  together with the expression of ${\cal R}'$ coming from (\ref{eqR}).
Hence, in this description, while the canonical field is the curvature 
fluctuations on comoving spatial hypersurfaces, the canonical momentum is 
the density contrast on the same hypersurfaces.

In order to bring the second-order action in the simple form  (\ref{Raction})
various (non-covariant) total derivatives have been dropped \cite{muk}. Hence, there is always 
 the freedom of redefining  the canonical fields through time-dependent functions of the background geometry. 
In particular, introducing 
\begin{equation}
v = - z {\cal R} = a \chi + z \Psi, 
\end{equation}
the following action can be obtained 
\begin{equation}
S^{(2)} = \frac{1}{2} \int d^4 x \biggl[ {v'}^2 -2 \frac{z'}{z} v v' - (\partial_{i} v)^2 + \biggl(\frac{z'}{z}\biggr)^2 v^2\biggr],
\label{sca2}
\end{equation}
whose related Hamiltonian and canonical momentum are, respectively
\begin{equation}
H^{(2)}(\eta) = \frac{1}{2} \int d^{3} x \biggl[ \pi_{v}^2 + 2 \pi_{v} v + (\partial_{i} v)^2\biggr],~~~~{\rm and}~
\pi = v' - \frac{z'}{z} v.
\label{hamscal2}
\end{equation}
In Eq. (\ref{sca2}) a further total derivative term can be dropped, leading to another action:
\begin{equation}
S^{(3)} = \frac{1}{2} \int d^4 x \biggl[ {v'}^2 - (\partial_{i} v)^2 +\frac{z''}{z} v^2\biggr],
\label{sca3}
\end{equation}
and another Hamiltonian
\begin{equation}
H^{(3)}(\eta) = \frac{1}{2} \int d^{3} x \biggl[ \tilde{\pi}_{v}^2 +  (\partial_{i} v)^2 - \frac{z''}{z} v^2\biggr].
\label{hamscal3}
\end{equation}
where $\tilde{\pi} = v'$. The Hamiltonians obtained in Eqs. (\ref{hamscal1}), (\ref{hamscal2}) and (\ref{hamscal3})
are related by canonical transformation and an example is provided in the Appendix, where 
a swift derivation of the analogous Hamiltonians is presented in the case of the tensor modes of the geometry.
Thus  Hamilton's  equations derived from the  Hamiltonians (\ref{hamscal1}), (\ref{hamscal2} and (\ref{hamscal3})
 will all have 
the same dynamical content. However, in spite of the dynamical equivalence of the descriptions, the 
quantum-mechanical states minimizing the different Hamiltonians will be different.

\renewcommand{\theequation}{3.\arabic{equation}}
\setcounter{equation}{0}
\section{Power spectra of curvature fluctuations}
In the present section the main  ``observable'' to be computed is the two-point function of curvature 
fluctuations for different spatial points but at the same time. This 
calculation will be consistently done in the case of the three 
examples discussed in the previous section, i.e.  (\ref{hamscal1}), (\ref{hamscal2}) and (\ref{hamscal3}).
The quantum mechanical-normalization will be imposed at {\em the same finite} value of the 
conformal time $\eta_0$.

Consider the situation when a  given quantum 
mechanical fluctuation is inside the horizon at a given time 
$\eta_0$. Suppose then to give initial conditions 
for the evolution of the quantum mechanical operators at
$\eta_0$ and adopt, for concreteness, the Heisenberg 
description. The  condition
\begin{equation}
k/a(\eta_{0}) = \Lambda,
\label{cond}
\end{equation}
defines the time $\eta_0$ at which a given physical scale crosses the NPH, i.e. the time at which 
the quantum mechanical initial conditions are assigned.
In order to make the  
calculations explicit, the case of exponential potentials 
\begin{eqnarray}
&& V = V_0 e^{ - \sqrt{\frac{2}{p}} \frac{ \varphi}{M_{\rm P}}},~~~~~~~~ 
\dot{\varphi}= \frac{ \sqrt{2 p} M_{\rm P}}{t}, 
\nonumber\\
&&z(t) = \sqrt{ \frac{2}{p}} M_{\rm P} a(t)
\label{pot}
\end{eqnarray}
will be studied. In Eqs. (\ref{pot}) the dot denotes a derivation with respect to the cosmic time $t$.

This set-up is sufficiently general to illustrate the different corrections arising 
in the case of different Hamiltonians. In particular notice that the slow-roll 
parameters\footnote{We denoted with $\epsilon$ and $\sigma$ the parameters related to the slow-roll
of the curvature and of the scalar field $\varphi$. Usually $\sigma$ is 
denoted by $\eta$, which would have generated confusion since 
this  letter is already  used, in the present notation, for the conformal time coordinate.}
\begin{equation}
\epsilon = \frac{M_{\rm P}^2}{2} \biggl(\frac{\partial\ln{ V}}{\partial\varphi}\biggr)^2 , 
~~~~~~~~~~~~\sigma =   -\frac{M_{\rm P}^2}{2} \biggl(\frac{\partial\ln{ V}}{\partial\varphi}\biggr)^2
+ \frac{M_{\rm P}^2 }{V} \frac{\partial^2 V}{\partial\varphi^2}
\label{sr}
\end{equation}
are all equal, i.e. $\epsilon = \sigma = 1/p$. This case can easily be  generalized to the 
situation where the slow-roll parameters are different (as in the case of polynomial potentials).

\subsection{The Hamiltonian for gauge-invariant curvature fluctuations}

In this case  the canonical field is ${\cal R}$, i.e. 
the curvature perturbation of Eq. (\ref{defR}). The canonical momentum is the density 
contrast as discussed in Eq. (\ref{piR}). The relevant Hamiltonian is given by
(\ref{hamscal1}).
The classical fields ${\cal R}$ and $\pi_{\cal R}$ can now be promoted to 
quantum-mechanical operators, obeying equal-time commutation relations 
\begin{equation}
[ \hat{\cal R}(\vec{x},\eta), \hat{\pi}_{\cal R}(\vec{y},\eta)] = 
i \delta^{(3)} (\vec{x} - \vec{y}),
\label{cancomm}
\end{equation}
so that the Hamiltonian operator will be 
\begin{equation}
\hat{H}(\eta) = \frac{1}{2} \int d^3 x 
\biggl[ \frac{\hat{\pi}_{\cal R}^2}{z^2} + z^2 (\partial_{i} \hat{\cal R})^2\biggr].
\label{ham1a}
\end{equation}
Going to Fourier space 
\begin{eqnarray}
&&\hat{{\cal R}}(\vec{x},\eta) = \frac{1}{2 (2\pi)^{3/2} } \int d^3 k \biggl[ \hat{{\cal R}}_{\vk} e^{- i \vec{k} \cdot \vec{x} }
+ \hat{{\cal R}}_{\vk}^{\dagger}  e^{ i \vec{k} \cdot \vec{x} }\biggr],
\nonumber\\
&& \hat{\pi}_{{\cal R}}(\vec{x},\eta) = \frac{1}{2 (2\pi)^{3/2} } \int d^3 k \biggl[ \hat{\pi}_{\vk} e^{- i \vec{k} \cdot \vec{x} }
+ \hat{\pi}_{\vk}^{\dagger}  e^{ i \vec{k} \cdot \vec{x} }\biggr],
\label{expansion}
\end{eqnarray}
the Hamiltonian will have the form
\begin{equation}
\hat{H}(\eta) = \frac{1}{4  } \int d^3 k \biggl[\frac{1}{z^2}(\hat{\pi}_{\vk} \hat{\pi}^{\dagger}_{\vk} + 
\hat{\pi}_{\vk}^{\dagger} \hat{\pi}_{\vk}) + k^2 z^2  (\hat{{\cal R}}_{\vk} \hat{{\cal R}}^{\dagger}_{\vk} + 
\hat{{\cal R}}_{\vk}^{\dagger} \hat{{\cal R}}_{\vk}) \biggr].
\label{ham1b}
\end{equation} 
Defining
\begin{equation}
\hat{Q}_{\vk}(\eta_0) = \frac{1}{\sqrt{2 k}} \biggl[\frac{\hat{\pi}_{\vk}(\eta_0)}{z(\eta_0)} -
i z(\eta_0) k \hat{{\cal R}}_{\vk}(\eta_0) \biggr],
\label{Q1}
\end{equation}
Eq. (\ref{ham1b}) can be  expressed as  
\begin{equation}
\hat{H}(\eta_0) =\frac{1}{4}\int d^{3} k k\biggl[ \hat{Q}^{\dagger}_{\vk} \hat{Q}_{\vk} 
+ \hat{Q}_{\vk} \hat{Q}^{\dagger}_{\vk} +\hat{Q}^{\dagger}_{-\vk} \hat{Q}_{-\vk} 
+ \hat{Q}_{-\vk} \hat{Q}^{\dagger}_{-\vk} \biggr],
\label{min1}
\end{equation}
while
canonical commutation relations between
conjugate field operators imply
 $[\hat{Q}_{\vk}, \hat{Q}_{\vp}^{\dagger} ] = \delta^{(3)}(\vec{k} - \vec{p})$.
Consequently, the state minimizing (\ref{ham1b}) at $\eta_0$ is the one annihilated by $\hat{Q}_{\vk}$, i.e. 
\begin{equation}
\hat{Q}_{\vk}(\eta_0) |0^{(1)}\rangle =0,~~~~~~~~~\hat{Q}_{-\vk}(\eta_0) |0^{(1)}\rangle =0.
\label{inst}
\end{equation}
The specific relation between field operators dictated by (\ref{inst}) provides 
initial conditions for the Heisenberg equations
\begin{equation}
i \hat{\cal R}' = [\hat{\cal R},\hat{H}],~~~~~~~~~~~~~
i \hat{\pi}_{\cal R}' = [\hat{\pi}_{\cal R},\hat{H}].
\label{hev}
\end{equation}
The full solution of this equation can be written as 
\begin{eqnarray}
&& \hat{\cal R}_{\vk}(\eta) = \hat{a}_{\vk}(\eta_0) f_{k}(\eta) + \hat{a}_{-\vk}^{\dagger}(\eta_0) f^{\ast}_{k}(\eta),
\label{solmu}\\
&& \hat{\pi}_{\vk}(\eta) = \hat{a}_{\vk}(\eta_0) g_{k}(\eta) + \hat{a}_{-\vk}^{\dagger}(\eta_0)g^{\ast}_{k}(\eta),
\label{solpi}
\end{eqnarray}
where, recalling the explicit solution of the equations in the case of the 
exponential potential (\ref{pot}) and defining $ x = k \eta$ 
\begin{eqnarray}
f_{k}(\eta) &=& \frac{\sqrt{\pi}}{4}\frac{e^{\frac{i}{2}(\mu + 1/2)\pi}}{z(\eta)\sqrt{ k}} 
\sqrt{-x} H^{(1)}_{\nu}(- x), ~~~~~~~~~~\nu= \frac{3 p -1}{2 (p -1)} 
\nonumber\\
g_{k}(\eta) &=& -\frac{\sqrt{\pi}}{4}e^{\frac{i}{2}(\mu + 1/2)\pi} z(\eta)\sqrt{k} \sqrt{- x} H^{(1)}_{\nu -1}(-x),
\label{fkgk}
\end{eqnarray}
satisfy the Wronskian normalization condition 
\begin{equation}
f_{k}(\eta) g^{\star}_{k}(\eta) - f^{\star}_{k}(\eta) g_{k}(\eta) = i.
\label{wrcon}
\end{equation}
The creation and annihilation operators appearing in (\ref{solpi}) are defined as 
\begin{eqnarray}
\hat{a}_{\vk}(\eta_{0}) &=& \frac{1}{z_0\sqrt{2 k}}  \{ [ g_{k}^{\ast}(\eta_{0}) 
+ i k z_0^2 f_{k}^{\ast}(\eta_0)] \hat{Q}_{\vk}(\eta_0) 
- [ g_{k}^{\ast}(\eta_{0} ) - i k z_0^2 f_{k}^{\ast}(\eta_0)]  \hat{Q}^{\dagger}_{-\vk}(\eta_0),
\nonumber\\
\hat{a}_{-\vk}^{\dagger}(\eta_{0}) &=& \frac{1}{z_0\sqrt{2 k}} \{ [ g_{k}(\eta_{0}) 
- i k z_0^2 f_{k}(\eta_0)] \hat{Q}^{\dagger}_{-\vk}(\eta_0)
- [ g_{k}(\eta_{0} ) + i k z_0^2 f_{k}(\eta_0)]  \hat{Q}_{\vk}(\eta_0).  
\label{def}
\end{eqnarray}
So far two sets of creation and annihilation operators have been introduced: the operators $\hat{Q}_{\vk}(\eta_0)$ and the 
operators $\hat{a}_{\vk}$. The state annihilated by $\hat{Q}_{\vk}(\eta_0)$ minimizes 
the Hamiltonian at $\eta_0$ while the state annihilated by $\hat{a}(\eta_0)$ {\em does not} minimize the Hamiltonian at 
$\eta_0$. The state annihilated by $\hat{a}_{k}(\eta_0)$ is the result of the unitary evolution of the fluctuations 
from $\eta= -\infty$ up to $\eta_0$. It is relevant to introduce these operators 
not so much for the calculation of the two-point function 
but for the subsequent applications to the back-reaction effects. In fact,
in the standard approach to the initial value problem for the 
quantum mechanical fluctuations, the initial state is chosen to be the one annihilated by $\hat{a}_{\vk}(\eta_0)$ for 
$\eta_{0} \to -\infty$. 

The Fourier transform 
of the two-point function, 
\begin{equation}
\langle 0^{(1)},\eta_0| \hat{{\cal R}}(\vx, \eta) \hat{\cal R}(\vy, \eta)| \eta_0, 0^{(1)} \rangle = 
\int \frac{d k}{k} {\cal P}_{{\cal R} }  \frac{\sin{ k r}}{k r},~~~~~~~~~~~~~r = |\vx - \vy|,
\label{PSA}
\end{equation}
can now be computed, and the result is   
\begin{eqnarray}
&&{\cal P}_{\cal R} = \frac{k^2}{2 \pi^2} 
\Biggl\{ |f_{k}(\eta)|^2 \biggl[ \frac{|g_{k}(\eta_0)|^2}{ z(\eta_0)^2} + k^2 z(\eta_0)^2 | f_{k}(\eta_0)|^2\biggr]
- \frac{f_{k}(\eta)^2}{2} \biggl[ \frac{{g_{k}^{\ast}(\eta_0)}^2 }{z(\eta_0)^2} + k^2 z(\eta_{0})^2 {f_{k}^{\ast}(\eta_0)}^2\biggr]
\nonumber\\
&& - \frac{{f_{k}^{\ast}(\eta)}^2}{2} \biggl[ \frac{{g_{k}(\eta_0)}^2 }{z(\eta_0)^2} + k^2 z(\eta_{0})^2 {f_{k}(\eta_0)}^2\biggr]\Biggr\}.
\label{ps1}
\end{eqnarray}
To derive Eq. (\ref{ps1}) it is useful to recall that, from (\ref{def}):
\begin{eqnarray}
&& \langle \eta_{0}, 0^{(1)} | \hat{a}_{\vk}(\eta_0) \hat{a}_{\vp}(\eta_0) | 0^{(1)}, \eta_0 \rangle = 
- \biggl[ \frac{z(\eta_0)^2 k}{2} {f_{k}^{\ast}}(\eta_0)^2 
+ \frac{1}{2 \, k\, z(\eta_{0})^2}{g_{k}^{\ast}}(\eta_0)^2 \biggr] \delta^{(3)} (\vk + \vp),
\nonumber\\
&& \langle \eta_{0}, 0^{(1)} | \hat{a}_{\vk}^{\dagger}(\eta_0) \hat{a}_{\vp}(\eta_0) | 0^{(1)}, \eta_0 \rangle 
= \biggl[\frac{k z(\eta_0)^2 }{2} |f_{k}(\eta_0)|^2  +
\frac{1}{2\,k\, z(\eta_0)^2} |g_{k}(\eta_0)|^2
\nonumber\\
&&- \frac{i}{2} ( f_{k}(\eta_0)^{\ast} g_{p}(\eta_{0}) - g_{k}(\eta_0)^{\ast} f_{p}(\eta_{0}))\biggr] \delta^{(3)} (\vk -\vp),
\end{eqnarray}
and similarly for the correlators of the Hermitian conjugates operator products.

To make contact with the  standard notation, scalar and tensor power spectra can be written as 
\begin{eqnarray}
&& {\cal P}_{\cal R} = \frac{25}{4} A_{S}^2,
\nonumber\\
&& {\cal P}_{h} = 25 A_{T}^2. 
\label{defps}
\end{eqnarray}
The explicit form of $A_{S}$ and $A_{T}$  can be obtained by inserting Eqs. (\ref{fkgk}) into Eq. (\ref{ps1}). 
The results should be expanded for  $|x| = k\eta \ll 1$ 
and  for $ |x_0| = k\eta_0 \gg 1 $.
While $|k\eta|$ measures how much a given mode is outside the horizon,
\begin{equation}
|x_0| = |k\eta_0| \simeq \frac{\Lambda}{H(t_{0}(k))} = \frac{\Lambda}{H_{\rm ex}}
\label{hex}
\end{equation}
defines the moment at which the given mode exits \footnote{ In this equation we denoted by $H(t_{0}(k))$ the moment 
at which a given mode exits   the NPH }.
In more explicit terms, the following relation holds  
\begin{equation}
\eta_{0}(k) = - \eta_1\biggl(\frac{\Lambda}{k} \biggr)^{1 - \frac{1}{p}}.
\end{equation}
As already pointed out, within the present approach, the time at which the quantum-mechanical normalization 
is implemented,  depends on the  comoving  wavelength. In the case of ordinary inflationary models
the initial time $\eta_{0} (\lambda)$  is directly proportional to a (positive) power of the 
comoving wavelength. This means that, for the same inflationary curvature scale, 
larger comoving wavelengths exit the NPH
earlier than  the small ones. 

The final result of the double expansion  then is 
\begin{eqnarray}
A_{S} &=& \sqrt{p}~{\cal C}(p) \biggl(\frac{H_1}{M_{\rm P}}\biggr)  \biggl( \frac{k}{k_1}\biggr)^{\frac{1}{p -1}}
\biggl[ 1 + \frac{p}{2 (p -1)} \frac{\sin{[ 2 x_0 + p \pi/(p-1)]}}{x_0}\biggr],
\nonumber\\
{\cal C}(p) &=&\frac{1}{5 \sqrt{2}} \frac{ 2^{\frac{p}{p-1}}}{\pi^{3/2} }
\biggl(\frac{p}{p -1}\biggr)^{-\frac{p}{p-1}} \Gamma\biggl( \frac{ 3 p -1}{2 (p-1)}\biggr)
\label{scps}\\
A_{T} &=& A_{S}/\sqrt{p}.
\label{tps}
\end{eqnarray}
The result of Eq. (\ref{tps}) (valid for the tensor modes of the geometry) is obtained 
by comparing Eq. (\ref{scps}) with the result for the 
correlator of the tensor fluctuations (see \cite{max1}). In (\ref{scps}) $H_1$ is the Hubble parameter 
at the end of inflation and $k_1 = H_1$ is the corresponding comoving 
frequency. Since, in the present case, $\epsilon(\varphi) = \sigma(\varphi) = 1/p$ (see Eq. (\ref{sr},
 Eqs. (\ref{scps}) and (\ref{tps}) imply $A_{T} = \sqrt{\epsilon} A_{S}$, which is the usual 
relation. 

In  Eqs. (\ref{scps}) and (\ref{tps}), on top of the standard (leading) terms
 there is a correction that goes, roughly, as 
$1/x_0 \sim H_{\rm ex}/\Lambda$ where, as discussed in Eq. (\ref{hex}), 
$H_{\rm ex}$ denotes the Hubble parameter evaluated at the 
moment the given scale exits the NPH. If $\Lambda \sim M_{\rm P}$, $H_{\rm ex}/\Lambda \sim 10^{-6}$. 
This is the correction that would apply 
in the scalar power spectrum if quantum mechanical initial conditions were assigned in such 
a way that the initial state minimizes (\ref{hamscal1}). As  will be shown in a moment, if a different Hamiltonian 
is minimized, the correction will be much smaller. 

\subsection{Hamiltonians for the canonical variable}
Having discussed in detail the results for the case of (\ref{hamscal1}) 
the attention will now be  turned to the case of (\ref{hamscal2}).
Following the same procedure discussed as previous case, commutation relations are imposed 
for the canonically conjugate fields 
\begin{equation}
[\hat{v}(\vx,\eta),\hat{\pi}_{v}(\vy,\eta) ] = i \delta^{(3)}(\vx - \vy),
\end{equation}
and the resulting Hamiltonian becomes \footnote{ The Fourier transform of the momentum operator $\hat{\pi}_{v}$ will be denoted, again, 
by $\hat{\pi}_{\vk}$.}: 
\begin{equation}
\hat{H}(\eta) = \frac{1}{4} \int d^3 k \biggl[(\hat{\pi}_{\vk} \hat{\pi}^{\dagger}_{\vk} + 
\hat{\pi}_{\vk}^{\dagger} \hat{\pi}_{\vk}) + k^2 (\hat{v}_{\vk} \hat{v}^{\dagger}_{\vk} + 
\hat{v}_{\vk}^{\dagger} \hat{v}_{\vk}) + \frac{z'}{z}( \hat{\pi}_{\vk} \hat{v}_{\vk}^{\dagger}+ 
\hat{\pi}_{\vk}^{\dagger} \hat{v}_{\vk} +  \hat{v}_{\vk} \hat{\pi}_{\vk}^{\dagger}+ 
\hat{v}_{\vk}^{\dagger} \hat{\pi}_{\vk})\biggr].
\label{ham2b}
\end{equation} 
Solving the evolution in the Heisenberg picture,  the mode functions can be written as 
\begin{eqnarray}
f_{k}(\eta) &=& \frac{\sqrt{\pi}}{4}\frac{e^{\frac{i}{2}( \nu + \frac{1}{2})\pi} }{\sqrt{ k}} \sqrt{-x} H^{(1)}_{\nu}(- x), 
\label{fk1}\\
g_{k}(\eta) &=&  -e^{\frac{i}{2}( \nu + \frac{1}{2})\pi} \frac{\sqrt{\pi}}{4}  \sqrt{k} \sqrt{-x} H^{(1)}_{\nu -1} (-x),
\label{gk1} 
\end{eqnarray}
where, as in the previous case, $2\nu = (3 p -1)/(p -1)$.
The quantum-mechanical state 
minimizing  (\ref{ham2b}) 
 at the initial time $\eta_{0}$, i.e. $|0^{(2)} \rangle $, is the one annihilated by $\hat{Q}_{k}$ whose definition is now
\begin{eqnarray}
\hat{Q}_{\vk}(\eta_0) = \frac{1}{\sqrt{2 k}} \biggl[e^{ -i \alpha_0 } \hat{\pi}_{\vk}(\eta_0) -
i e^{  i \alpha_0 } k \hat{v}_{\vk}(\eta_0) \biggr],
\nonumber\\
\hat{Q}_{\vk} |0^{(2)} \rangle =0, ~~~~~\hat{Q}_{-\vk} |0^{(2)} \rangle =0,
\label{Q}
\end{eqnarray}
where $\alpha_0 = \alpha(\eta_0)$, i.e. 
\begin{equation}
\sin{2\alpha_0} =\left. \frac{z'}{k z}\right|_{\eta_0}.
\end{equation} 
The canonical commutation relations Eq. (\ref{cancomm}) now imply  
$[\hat{Q}_{\vk}, \hat{Q}_{\vp}^{\dagger}] = 
\cos{2\alpha } \delta^{(3)}(\vec{k} -\vec{p})$. 
In terms of (\ref{Q}) the Hamiltonian (\ref{ham2b}) 
has again the form (\ref{min1}) but, obviously, the meaning of $\hat{Q}_{\vk}$ is different.

The wave-functional 
of the initial state  has a Gaussian form:
\begin{equation}
\psi [ v_{\vk}] = N {\rm exp}
 \left( - \sum_k \frac{k}{2} (v_{\vk} v_{-\vk}) e^{-2i\alpha_0} \right).
\label{wf}
\end{equation}
This state is  normalizable provided $|\alpha_0 |< \pi/4$. 
We see that $|\alpha_0 | = \pi/4$
corresponds to a time $\eta_0$ for which $|k\eta_0| \sim 1$ (recall that $z'/z $ goes as $1/\eta$), 
which is basically equivalent 
to the condition of (standard) horizon crossing. Consequently,  provided the modes of the field 
are inside the horizon at the ``initial" time $\eta_0$, the state (\ref{wf}) is normalizable.

The two-point function of the curvature fluctuations can now be computed, with the difference that, now,
the state minimizing (\ref{ham2b}) at $\eta_0$ is the one 
annihilated by (\ref{Q}), i.e.
\begin{equation}
\langle 0^{(2)},\eta_0| \hat{{\cal R}}(\vx, \eta) \hat{\cal R}(\vy, \eta)| \eta_0, 0^{(2)} \rangle = 
\int \frac{d k}{k} {\cal P}_{{\cal R} }  \frac{\sin{ k r}}{k r},~~~~~~~~~~~~~r = |\vx - \vy|,
\label{PSA1}
\end{equation}
 The result of this calculation 
follows the same steps as outlined before and, recalling the definitions (\ref{defps}), the results are
\begin{eqnarray}
A_{S} &=& \sqrt{p}~ {\cal C}(p) \biggl(\frac{H_1}{M_{\rm P}}\biggr)  \biggl( \frac{k}{k_1}\biggr)^{\frac{1}{p -1}}
\biggl[ 1 -\frac{p}{4 (p -1)} \frac{\cos{[ 2 x_0 + p \pi/(p-1)]}}{x_0^2}\biggr],
\nonumber\\
A_{T} &=& A_{S}/\sqrt{p},
\label{tps2}
\end{eqnarray}
where ${\cal C}(p)$ is the same as in (\ref{scps}). A comparison of Eqs. (\ref{tps}) and (\ref{tps2}) shows two important 
facts. The first is that the leading term of the spectrum is {\em the same in both cases}. Furthermore, it will be shown that
this is true even if (\ref{hamscal3}) is used. This phenomenon simply reflects the 
fact that different Hamiltonians, connected by canonical transformations, must lead to the same evolution 
and to the same leading term in the power spectra. The second fact to be noticed is that the correction 
to the power spectrum goes as $1/x_0^2$ in the case of (\ref{tps2}). This correction is then much smaller than 
the one appearing in (\ref{tps}). If $\Lambda \sim M_{\rm P}$ then the correction will be ${\cal O}(10^{-12})$, i.e.
six orders of magnitude smaller than in the case of (\ref{tps}).

Finally the case of  the  Hamiltonian (\ref{hamscal3}) will be examined.
Equation (\ref{hamscal3})  can be minimized following the same procedure as
already discussed in the case of Eqs. (\ref{hamscal1}) and (\ref{hamscal2}). 
Defining the function 
\begin{eqnarray}
\omega^2(x) = \biggl( 1 - \frac{1}{k^2} \frac{a''}{a}\biggr),
\end{eqnarray}
and recalling that $\tilde{\pi}_{v} = v'$,
the Hamiltonian (\ref{ham3a}) can be written in the simple form 
\begin{equation}
\hat{H}(\eta) = \frac{1}{4}\int d^{3} k \biggl[(\hat{\tilde{\pi}}_{\vk} \hat{\tilde{\pi}}^{\dagger}_{\vk} + 
\hat{\tilde{\pi}}_{\vk}^{\dagger} \hat{\tilde{\pi}}_{\vk}) + k^2\omega^2(x) (\hat{v}_{\vk} \hat{v}^{\dagger}_{\vk} + 
\hat{v}_{\vk}^{\dagger} \hat{v}_{\vk})\biggr].
\label{ham3a}
\end{equation}
Defining now the  operator
\begin{equation}
\hat{Q}_{\vk}(\eta_0) = \frac{1}{\sqrt{2 k}}\biggl[ \hat{\tilde{\pi}}_{\vk}(\eta_0) - i k  \omega(\eta_0) 
\hat{v}_{\vk}(\eta_0)\biggr],
\label{Qpr}
\end{equation}
the Hamiltonian can  again be expressed, at $\eta_0$ in the same  form as
previously discussed, namely, the one given by Eq. (\ref{min1}), with the caveat that 
now the operator (\ref{Qpr}), if compared with that  defined in Eq. (\ref{Q}),
 has a different expression in terms of the 
canonical fields.
The commutation relations now are
 $[\hat{Q}_{\vk}(\eta_0), \hat{Q}_{\vp}^{\dagger}(\eta_0)] = \omega_0 \delta^{(3)}(\vec{k}-\vec{p})$.
The mode functions
  $f_{k}(\eta)$ are the same as those given in Eq. (\ref{fk1}), while  ${g}_{k}$ is given by
\begin{eqnarray}
g_{k}(\eta) &=&  - \frac{\sqrt{\pi}}{4} e^{\frac{i}{2}( \nu + \frac{1}{2})\pi}\sqrt{k} \sqrt{-x}\biggl[ H^{(1)}_{\nu -1} (-x) +
\frac{(1 -2 \nu)}{2(-x)} H^{(1)}_{\nu} (-x)\biggr] ,
\label{tgk} 
\end{eqnarray}
Repeating the steps used in the previous two cases the two-point function
\begin{equation}
\langle 0^{(3)},\eta_0| \hat{{\cal R}}(\vx, \eta) \hat{\cal R}(\vy, \eta)| \eta_0, 0^{(3)} \rangle
\end{equation}
can be computed recalling that $|0^{(3)}\rangle$ is the state annihilated by (\ref{Qpr}).
The following power spectra are then obtained
\begin{eqnarray}
A_{S} &=& \sqrt{p}~{\cal C}(p) \biggl(\frac{H_1}{M_{\rm P}}\biggr)  \biggl( \frac{k}{k_1}\biggr)^{\frac{1}{p -1}}
\biggl[ 1 + \frac{p( 2 p-1)}{(p -1)} \frac{\sin{[ 2 x_0 + p \pi/(p-1)]}}{4 x_0^3}\biggr],
\label{scps3}\\
A_{T} &=& A_{S}/\sqrt{p}.
\label{tps3}
\end{eqnarray}
In Eqs. (\ref{scps3}) and (\ref{tps3}) 
 the correction arising from the initial state goes as $1/x_0^3$ and, again, if $\Lambda \sim M_{\rm P}$ 
it is ${\cal O}(10^{-18})$, i.e. $12$ orders of magnitude smaller than in the case discussed 
in Eqs. (\ref{scps}) and (\ref{tps}).

\renewcommand{\theequation}{4.\arabic{equation}}
\setcounter{equation}{0}
\section{Resolving the ambiguity: back-reaction effects} 

In order to select the correct Hamiltonian in a way compatible with the 
idea of assigning initial conditions on the NPH, it is desirable  to 
address the issue of back-reaction effects. 
The energetic content of the quantum-mechanical state minimizing 
the given Hamiltonian should be estimated and  compared with the energy density of the background geometry.
The back-reaction 
effects of the different quantum-mechanical states 
minimizing the Hamiltonians will now be computed. 
Without loss of generality, the attention will 
be focused on the tensor modes of the geometry. In the appendix it is shown, 
the same Hamiltonians as were discussed for the scalar modes of the geometry can also be defined 
in the case of the tensors. Moreover, there is one-to-one correspondence between scalar and tensor Hamiltonians. 
The advantage of discussing the gravitons is that they do not couple to 
the sources and, therefore, the form of the energy-momentum pseudo-tensor is simpler 
than in the case of the scalar modes \cite{br3}.

The appropriate energy-momentum tensor of the fluctuations 
of the geometry will be averaged over the state 
minimizing a given Hamiltonian at $\eta_0$ and the result 
compared with the energy density  of the background geometry. 
The energy density of the fluctuations cannot exceed 
that of the background geometry. 

The energy density of the gravitational waves can be computed from 
the energy-momentum pseudo-tensor \cite{br3}, written, for simplicity, for one 
of the two traceless and divergenceless polarizations (i.e. $h_{i}^{i}=0$ and $\partial_{i} h^{i}_{j} =0$) :
\begin{equation}
\langle \hat{{\cal T}}_{0}^{0} \rangle = 
\frac{{\cal H}}{2 a^2} \langle (\hat{h}' \hat{h} + \hat{h} \hat{h}') \rangle  
+ \frac{1}{8 a^2} \langle [ \hat{h}'^2  + (\partial_{i} \hat{h})^2 ] \rangle, 
\end{equation}
where $\langle ...\rangle$ denotes the expectation value with respect 
to a quantum mechanical state minimizing a given Hamiltonian and $\hat{h}$ denotes the field 
operator corresponding to a single tensor 
polarization of the geometry. 

In the appendix  we swiftly recall the main quantities that are required 
in order to discuss the minimization of the Hamiltonians of the tensor 
modes of the geometry. This notation will be followed here too.
Consider, to begin with, the first Hamiltonian, i.e. (\ref{H1t}). 
The  quantity 
\begin{equation}
\rho_{\rm GW}^{(1)}(\eta,\eta_{0}) =
 \langle 0^{(1)}, \eta_{0}| \hat{{\cal T}}_{0}^{0}(\eta) | \eta_{0}, 0^{(1)} \rangle, 
\end{equation}
should be computed. Here $|\eta_{0}, 0^{(1)}\rangle$ is the state minimizing the first Hamiltonian, i.e. 
the state annihilated by 
\begin{equation}
\hat{Q}_{\vk}(\eta_0) = \frac{1}{\sqrt{2 k}} \biggl[\frac{\hat{\Pi}_{\vk}(\eta_0)}{a(\eta_0)} -
i a(\eta_0) k \hat{h}_{\vk}(\eta_0) \biggr].
\label{Q1t}
\end{equation}
As illustrated in the case of the scalar modes of the geometry, the 
evolution equations in the Heisenberg description can be 
solved in terms on the values of the field operators 
at the initial time $\eta_0$. 
Using repeatedly the action of the operators (\ref{Q1t}), 
\begin{eqnarray}
\rho_{\rm GW}^{(1)}(\eta,\eta_{0}) &=& \frac{H^4}{32 \pi^2} \int \frac{d k}{k} x^4 
\biggl[ \frac{a^2(\eta)}{a^2(\eta_0)} A_{k}^{(1)}(\eta,\eta_0)^2 + \frac{a^2(\eta_0)}{a^2(\eta)} D_{k}^{(1)}(\eta,\eta_0)^2 
\nonumber\\
&+& 
\frac{C_{k}^{(1)}(\eta,\eta_0)^2}{k^2 a^2(\eta_0) a^2(\eta)} + k^2 a^2(\eta) a^2(\eta_0) B_{k}^{(1)}(\eta,\eta_0)^2 
\nonumber\\
&+&8 {\cal H} \frac{A_{k}^{(1)}(\eta,\eta_0)C_{k}^{(1)}(\eta,\eta_0)}{k^2 a^2(\eta_0)} 
+ 8 {\cal H} a(\eta_0)^2 D_{k}^{(1)}(\eta,\eta_0) B_{k}^{(1)}(\eta,\eta_0) \biggr],
\label{GWpt1}
\end{eqnarray}
where, defining $\Delta x = x - x_0$,
\begin{eqnarray}
&& A^{(1)}_{k}(\eta,\eta_0) =\frac{x\,\cos{\Delta x} - \sin{\Delta x}}{x_0} ,
\nonumber\\
&& B^{(1)}_{k}(\eta,\eta_0) = \frac{H^2\,\left[ \left( -x + x_0 \right) \,\cos{\Delta x} + 
      \left( 1 + x\,x_0 \right) \,\sin{\Delta x} \right] }{k^3},
\nonumber\\
&& C^{(1)}_{k}(\eta,\eta_0) =-\left( \frac{k^3\,\sin{\Delta x}}{H^2\,x\,x_0} \right) ,
\nonumber\\
&& D^{(1)}_{k}(\eta,\eta_0) = \frac{x_0\,\cos{\Delta x} + \sin{\Delta x}}{x}.
\label{C1}
\end{eqnarray}
Inserting Eqs. (\ref{C1}) into Eq. (\ref{GWpt1}) the following expression can be obtained:
\begin{eqnarray}
\rho_{\rm GW}^{(1)}(\eta,\eta_0) &=& \frac{H^4}{64 \pi^2} \int \frac{ d k}{k} \biggl(\frac{x}{x_0}\biggr)^2 
\biggl\{  ( 2 x_0^2 + 1) ( 2 x^2 - 7) + (12 x x_0 + 7) \cos{2 \Delta x}
\nonumber\\ 
&+& 2 ( 3 x - 7 x_0) \sin {2 \Delta x} \biggr\}.
\label{GWpt1a}
\end{eqnarray}
where the explicit paramertization of de Sitter space has been used, namely, 
$a(\eta) = (- H\eta)^{-1}$. 

In curved space-times, it is often mandatory to implement 
a suitable renormalization procedure, which amounts, in some cases, to subtracting the 
appropriate counter terms which can  in turn be expressed as known geometrical quantities.
Instead of looking immediately at this problem from a formal point of view, it is useful 
to see what physics suggests. Let us go back to the logic behind the present exercise. 
We want to assign quantum-mechanical initial conditions for the field operators at a 
{\em finite} value of the conformal time, as soon as the physical wavelength 
becomes sub-Planckian. Following this logic, the energy-density 
present for $\eta_0 \to - \infty$ has no meaning since, in this limit, all the 
physical wavelength will go to to $0$, i.e. will be much smaller than the Planck length.
Let us then take the limit of Eq. (\ref{GWpt1a}) for $\eta_0 \to -\infty$:
\begin{equation} 
\lim_{\eta_0 \to -\infty} \rho_{\rm GW}^{(1)} =  \frac{H^4}{64 \pi^2} \int \frac{ d k}{k} 2  x^2 ( 2 x^2  - 7) \equiv 
\langle \hat{{\cal T}}_{0}^{0} \rangle _{\rm BD}.
\label{BD1a}
\end{equation}
This quantity, as can be checked from the other expectation values 
presented later in this section, is the {\em same for all three Hamiltonians}. 
The result of Eq. (\ref{BD1a}) has a simple interpretation: it is the 
expectation value of the energy-momentum 
pseudo-tensor over the Bunch--Davies vacuum for $\eta \to -\infty$.
In fact $\langle \hat{\cal T}_{0}^{0} \rangle_{\rm BD} $ can be obtained  by averaging each of the Hamiltonians discussed in the 
present section over the state annihilated by the corresponding operators $\hat{a}_{\vk}$ and $\hat{a}_{-\vk}$. These operators
have been discussed in Eqs. (\ref{solmu}) and (\ref{solpi}) in the case of the first class of scalar Hamiltonians. There it was 
noticed that when $\eta_0$ is finite, the state annihilated by $\hat{a}_{\vk}$ and $\hat{a}_{-\vk}$
does not minimize any Hamiltonian. Clearly the same set of operators can be defined, with the appropriate differences, for
all the Hamiltonians discussed in the present investigation.
In \cite{tanaka} it was suggested 
that a sensible renormalization procedure amounts to subtracting the energy density of the 
Bunch-Davies vacuum. 

The renormalized energy density can then be defined as  
\begin{eqnarray}
\overline{\rho}_{\rm GW}^{(1)}(\eta,\eta_0) &=&   \rho_{\rm GW}^{(1)}(\eta,\eta_0) - \langle {\cal T}_{0}^{0} \rangle _{\rm BD}
= \frac{H^4}{64 \pi^2} \int \frac{ d k}{k} \biggl(\frac{x}{x_0}\biggr)^2 \biggl\{  
( 2 x_0^2 -7)
\nonumber\\
&+& (12 x x_0 + 7) \cos{2 \Delta x} + 2 ( 3 x - 7 x_0) \sin {2 \Delta x} \biggr\}.
\end{eqnarray}
Recall now that $x = k\eta$. Then
integrating between $|x| \sim 1$ and $ |x|\sim x_0$, and keeping only the leading terms for 
$x_0 \gg1$, we have the following result:
\begin{equation}
\overline{\rho}_{\rm GW}^{(1)}(\eta,\eta_0) \sim \frac{H^4}{64 \pi^2}\biggl[ x_0^2 + {\cal O} \biggl(\frac{1}{x_0^2}\biggr)\biggr]\simeq 
\frac{H^4}{64 \pi^2}\biggl(\frac{\Lambda}{H}\biggr)^2\biggl[ 1 + {\cal O}\biggl(\frac{H^2}{\Lambda^2}\biggr)\biggr].
\label{en1}
\end{equation}
It can be checked numerically that the agreement of (\ref{en1}) with the exact result of the integral 
is excellent.
Since, as already discussed, $ |x_0| = \Lambda/H \gg 1$, in the case of de Sitter space, the back-reaction effects 
related to the state minimizing the first Hamiltonian are then large. Recall, in fact, that the energy density of 
the background geometry is ${\cal O} ( H^2 M_{\rm P}^2)$. Hence, if $\Lambda \sim M_{\rm P}$ the energy 
density of the fluctuations will be of the same order as that  of the background geometry, which is 
not acceptable since, if this is the case, inflation could not even start.

Let us now turn our attention to the case of the state minimizing the second Hamiltonian,
The expectation value of the energy-momentum pseudo-tensor
\begin{eqnarray}
\rho_{\rm GW}^{(2)}(\eta,\eta_{0}) &=&
 \langle 0^{(2)}, \eta_{0}| \hat{{\cal T}}_{0}^{0}(\eta) | \eta_{0}, 0^{(2)} \rangle
\nonumber\\
&=& \frac{H^4}{64 \pi^2 } \int \frac{d k}{k} \frac{x^4}{\cos{2 \alpha_0}}
 \biggl\{ A_{k}^{(2)}(\eta,\eta_0)^2 + \frac{C_{k}^{(2)}(\eta,\eta_0)^2}{k^2} + 
8 {\cal H} D_{k}^{(2)} (\eta,\eta_0)  B_{k}^{(2)}(\eta,\eta_0) 
\nonumber\\
&+& 8 {\cal H} A_{k}^{(2)}(\eta,\eta_0) C_{k}^{(2)}(\eta,\eta_0)
+ D_{k}^{(2)}(\eta,\eta_0)^2  + k^2 B_{k}^{(2)}(\eta,\eta_0)^2  
\nonumber\\
&-& \frac{2}{k} \sin{2 \alpha_0} \biggl[ C_{k}^{(2)}(\eta,\eta_0) D_{k}^{(2)}(\eta,\eta_0) + 
k^2 B_{k}^{(2)}(\eta,\eta_0) A_{k}^{(2)}(\eta,\eta_0) 
\nonumber\\
&-& 4 {\cal H} \biggl( A_{k}^{(2)}(\eta,\eta_0) D_{k}^{(2)}(\eta,\eta_0) +  B_{k}^{(2)}(\eta,\eta_0) 
C_{k}^{(2)}(\eta,\eta_0)\biggr) \biggl]\biggl\}
\label{GWpt2}
\end{eqnarray}
is taken over the state minimizing the second Hamiltonian, i.e.  the state annihilated by 
\begin{equation}
 \hat{Q}_{\vk}(\eta_0) = \frac{1}{\sqrt{2 k}} \biggl[e^{ -i \alpha_0 } \hat{\pi}_{\vk}(\eta_0) -
i e^{  i \alpha_0 } k \hat{\mu}_{\vk}(\eta_0) \biggr].
\label{Q2t}
\end{equation}
Recall that, in Eq. (\ref{GWpt2}), $\sin{2 \alpha_0} = -1/x_0$ and ${\cal H}_{0} = -1/\eta_0$.
Taking into account the explicit form of the coefficients
\begin{eqnarray}
&& A_{k}^{(2)}(\eta,\eta_0) = \cos{\Delta x} - \frac{\sin{\Delta x}}{x},
\nonumber\\
&& B_{k}^{(2)}(\eta,\eta_0) = \frac{(1 + x x_0) \sin{\Delta x} - \Delta x \cos{\Delta x}}{ k x x_0 },
\nonumber\\
&& C_{k}^{(2)}(\eta, \eta_0) = - k \sin{\Delta x},
\nonumber\\
&& D_{k}^{(2)}(\eta,\eta_0) = \cos{\Delta x} + \frac{\sin{\Delta x}}{x_0}, 
\end{eqnarray}
the renormalized energy-momentum pseudo-tensor becomes:
\begin{eqnarray}
\overline{\rho}_{\rm GW}^{(2)}(\eta,\eta_0) &=& \frac{H^4}{64 \pi^2} \int \frac{d k}{k} 
\biggl(\frac{x}{x_0}\biggr)^2 \frac{1}{\cos{2\alpha_{0}}} \biggl\{ ( 2 x^2 - 7) [ 2 x_0^2 -1 - 2 x_0^2 \cos{ 2\alpha_{0}}] 
\nonumber\\
&-& 7 \cos{ 2 \Delta x} - 6 \sin{ 2 \Delta x} \biggl\}.
\end{eqnarray}
Applying the procedure described above and performing the integral over all the modes inside 
the horizon, but below the cut-off $\Lambda$, we have 
\begin{equation}
\overline{\rho}_{\rm GW}^{(2)}\eta,\eta_0) \simeq - \frac{25}{ 512 \pi^2}\, H^4\,
\biggl[ 1 + {\cal O}\biggl(\frac{H^2}{\Lambda^2}\biggr)\biggr].
\label{en2}
\end{equation}
In this case the energy density is smaller, by a factor of $(H/\Lambda)^2$, than the energy density obtained in (\ref{en1}).
If we took $ \Lambda \sim M_{\rm P}$, this result would be acceptable except for the sign of the averaged 
energy density, which is negative. The fact that negative energy densitiescould be obtained,
  by averaging the energy density over a specific 
quantum state   was noted long ago by Ford and Kuo \cite{ford3} (see also \cite{ford4}). 
In \cite{ford3} it was actually noticed that the averaged energy density becomes negative whenever the fluctuations of the 
energy momentum tensor itself are large. 

Finally, in the third and last case,  the state minimizing the Hamiltonian (\ref{H3t}) is the one 
annihilated by 
\begin{equation}
\hat{Q}_{\vk}(\eta_0) = \frac{1}{\sqrt{2 k}} \biggl[ \hat{\tilde{\pi}}_{\vk}(\eta_0) - i \omega_{0} \hat{\mu}_{\vk}(\eta_0) \biggr],
\end{equation}
where 
\begin{equation}
\omega_0 = \sqrt{ 1 - \frac{({\cal H}_0^2 + {\cal H}'_0)}{k^2}}= \sqrt{1 - \frac{2}{x_0^2}}.
\end{equation}
The average of the energy-momentum pseudo-tensor over this state 
leads to the following expression 
\begin{eqnarray}
\rho_{\rm GW}^{(3)}(\eta,\eta_0) &=& \langle 0^{(3)}, \eta_{0}| \hat{{\cal T}}_{0}^{0}(\eta) | \eta_{0}, 0^{(3)} \rangle
\nonumber\\
&=& \frac{H^4}{32\pi^2} \int \frac{d k}{k} \frac{x^4}{\omega_0} \Biggl\{ \biggl( 1 - 7 \frac{{\cal H}}{k^2} \biggr) \biggl[ 
A_{k}^{(3)}(\eta,\eta_0)^2 + k^2 \omega_0^2 B_{k}^{(3)}(\eta,\eta_0)^2\biggr] 
\nonumber\\
&+& \frac{6 {\cal H}}{k^2} \biggl[ A_{k}^{(3)}(\eta,\eta_0) C_{k}^{(3)}(\eta,\eta_0) 
+ k^2 \omega_0^2 B_{k}^{(3)}(\eta,\eta_0) D_{k}^{(3)}(\eta,\eta_0)\biggr]  
\nonumber\\
&+& \frac{1}{k^2}\biggl[C_{k}^{(3)}(\eta,\eta_0)^2 
+ k^2 \omega_0^2 D_{k}^{(3)}(\eta,\eta_0)^2\biggr] \Biggr\}.
\end{eqnarray}
Recalling that 
\begin{eqnarray}
A_{k}^{(3)}(\eta,\eta_0) &=& \frac{\left( x_0 +  x\,\left(  {x_0}^2 -1 \right)\right) \,\cos{\Delta x} + 
    \left( 1 + \left( x - x_0 \right) \, x_0 \right) \,\sin{\Delta x}}{x\,{x_0}^2},
\nonumber\\
B_{k}^{(3)}(\eta,\eta_0) &=&\frac{\left( -x + x_0 \right) \,\cos{\Delta x} + \left( 1 + 
       x\,x_0 \right) \,\sin{\Delta x}}{k\,x\,x_0},
\nonumber\\
C_{k}^{(3)}(\eta,\eta_0) &=& - \frac{k\left[ \left( x_0 - x^2\,x_0+ x\left(  
             {x_0}^2-1\right)  \right) \cos{\Delta x} +
         \left( 1 + x x_0- {x_0}^2 + x^2
            \left( {x_0}^2-1\right)  \right) \sin{\Delta x}\right] }{x^2\,{x_0}^2}, 
\nonumber\\
D_{k}^{(3)}(\eta,\eta_0) &=&\frac{\left( x - x_0+ x^2\,x_0\right) \,\cos{\Delta x} + 
    \left( -1 + x^2 -  x\,x_0\right) \,\sin{ \Delta x}}{x^2\,x_0},
\end{eqnarray}
we arrive at the following final expression
\begin{eqnarray}
\overline{\rho}_{\rm GW}^{(3)}(\eta,\eta_0) &=& \frac{H^4}{64 \pi^2} \int \frac{d k}{k} \frac{ x^2}{x_0^4 \omega_0} \biggl\{ 
( 2 x^2 -7) ( 2 x_0^4 - 2 x_0^2 -1 - 2 x_0^4 \omega_0) 
\nonumber\\
&-& ( 12 x x_0 + 7) \cos{2 \Delta x} + 2 (7 x_0 - 3 x) \sin{2 \Delta x} \biggr\}.
\end{eqnarray}
Applying the same procedure as  described above and integrating over $k$, we have 
\begin{equation}
\overline{\rho}_{\rm GW}^{(3)}(\eta,\eta_0) \simeq \frac{27}{256 \pi^2}\,H^4\, \biggl(\frac{H}{\Lambda}\biggr)^2\biggl[ 1  + {\cal O}
\biggl(\frac{H^4}{\Lambda ^4}\biggr)\biggr],
\label{en3}
\end{equation}
i.e. even smaller than the result discussed in the case of the second Hamiltonian. In this case, the averaged 
energy density is much smaller 
than that of the background, and it is positive.

Thus, by taking the average of the energy-momentum pseudo-tensor, we were able to give an intrinsic characterization 
of the back-reaction effects that arise when quantum-mechanical initial condition are assigned at a finite 
time and for all the physical wavelength  in excess of a given fundamental scale $\Lambda^{-1}$.
\renewcommand{\theequation}{6.\arabic{equation}}
\setcounter{equation}{0}
\section{Concluding remarks }
If quantum-mechanical initial conditions are assigned at a finite time $\eta_0$ during inflation 
and in excess of a 
given physical wavelength, ambiguities may arise.
The presence 
of a cut-off $\Lambda$ in the physical momenta $k/a(\eta)$, by itself, 
does not specify any corrections in the late-time observables. In order to predict quantitatively 
the corrections in the scalar and tensor power spectra, it is mandatory to 
give a prescription for assigning the quantum mechanical normalization of the fluctuations.
In general terms, the answer to this question is that quantum-mechanical fluctuations should minimize 
at $\eta_{0}$  a given Hamiltonian. However, since the problem is time-dependent, canonical 
tranformations can change the form of the Hamiltonian. Different Hamiltonians, related by canonical 
tranformations, lead to the same {\em evolution of the fluctuations}. However, since 
the initial state differs for each of the selected Hamiltonians, the power spectra of scalar and tensor modes 
will inherit computable corrections. Examples of this phenomenon were given in the present paper. 
Various, rather natural, Hamiltonians can be defined in the analysis of cosmological perturbations. It has been 
shown that the most adiabatic Hamiltonians lead to the smallest corrections, not only for the 
tensor modes of the geometry \cite{max1} but also for the scalar modes. 

The criteria used to select one prescription or the other cannot be 
 related only to ``achievable'' magnitude of the corrections in the two-point functions. On the contrary, it is important 
to estimate the energetic content of the initial states by  minimizing the different Hamiltonians. The energy 
density of the quantum fluctuations should then be compared with that
of the background geometry. This procedure provides a way of discarding initial states 
on the basis of excessive back-reaction effects. This exercise has been performed in detail, 
making use of the energy-momentum pseudo-tensor which is a common tool in the 
analysis of back-reaction effects of metric fluctuations.

If non-adiabatic Hamiltonians are minimized at $\eta_0$, the corrections to the tensor power
spectrum can be as large as $10^{-6}$. However, the energy density 
of the initial state is, in this case, comparable with the energy density 
of the background geometry if $ \Lambda \sim M_{\rm P}$. On the contrary, if 
adiabatic Hamiltonians are minimized at $\eta_0$, the energy density of the fluctuations 
is always smaller than that of the background geometry by $12$ or $24$ orders of magnitude 
making then negligible back-reactiuon effects.

\section*{Acknowledgements}

The author is indebted to G. Veneziano for very useful discussions
and encouragments. Interesting discussions with V. Bozza, I. Tkachev and R. 
Woodard are also acknowledged.

\newpage
\begin{appendix}

\renewcommand{\theequation}{A.\arabic{equation}}
\setcounter{equation}{0}
\section{Tensor modes of the geometry} 
In this appendix the explicit results concerning the tensor modes 
of the geometry will be swiftly recalled in view of the applications 
related to the back-reaction effects.
The quadratic action for the tensor modes of the geometry can be written 
as 
\begin{equation}
S_{\rm GW} = \frac{ 1}{64 \pi G} \int d^4 x a^2 \partial_{\alpha} h_{i}^{j} \partial_{\beta} h_{j}^{i} \eta^{\alpha\beta},
\end{equation}
where $\eta_{\alpha\beta}$ is the 
Minkowski metric and where $ \partial_{i}h^{i}_{j} = h_{i}^{i} =0$. In this 
gauge-invariant splitting of the degrees of freedom of 
the perturbed metric, the gravitational wave is 
a rank 2 tensor in three spatial dimensions, which is symmmetric, traceless and divergenceless. 
If we then consider the action of a single polarization and redefine, accordingly, the tensor amplitude 
in order to include the Planck mass, we will see that the action is given by 
\begin{equation}
S_{\rm GW}^{(1)} = \frac{1}{2} \int d^{4} a^2 \partial_{\alpha} h \partial_{\beta} h \eta^{\alpha\beta},
\end{equation}
whose canonical momentum is simply given by $\Pi= a^2 h'$ and whose associated 
Hamiltonian is 
\begin{equation}
H^{(1)}_{\rm GW}(\eta) =\frac{1}{2} \int d^3 x \biggl[ \frac{ \Pi^2}{ a^2 } 
+ a^2 (\partial_{i}h)^2\biggr].
\label{H1t}
\end{equation}
Since this Hamiltonian is time-dependent, it is always possible to perform 
time-dependent canonical transformations, leading to a different 
Hamiltonian. In particular, if we define the rescaled field, $\mu = a h$, the corresponding 
action will become 
\begin{equation}
S^{(2)}_{\rm GW} =\frac{1}{2} \int d^4 x \biggl[ {\mu '}^2 - 2 {\cal H} \mu \mu' 
+ {\cal H}^2 \mu^2 + (\partial_{i}\mu)^2 \biggr],
\end{equation}
while the associated Hamiltonian can be written as 
\begin{equation} 
H^{(2)}_{\rm GW}(\eta) = \int d^{3} x \biggl[  \pi^2 +  2{\cal H} \mu \pi +
 (\partial_{i} \mu)^2 
\biggr],
\label{H2t}
\end{equation}
in terms of $\mu$ and of the canonically 
conjugate momentum $\pi = \mu' - {\cal H} \mu$.
Finally, a further canonical transformation 
can be performed 
starting from (\ref{H2t}). Defining 
the generating functional in terms of the 
old fields $\mu$ and of the new momenta $\tilde{\pi}$,  
\begin{equation}
{\cal F}_{2\to 3} ( \mu, \tilde{\pi}, \eta) = \int d^{3} x \biggl( \mu \tilde{\pi},
 - \frac{{\cal H}}{2} \mu^2\biggr),
\label{2to3}
\end{equation} 
the new Hamiltonian can be obtained by taking the partial 
(time) derivative of (\ref{2to3}), with the result
\begin{equation}
H^{(3)}_{\rm gw}(\eta) = \frac{1}{2}\int d^{3} x \biggl[ \tilde{\pi}^2 + 
(\partial_{i} \mu)^2 - ( {\cal H}^2 + {\cal H}')  \mu^2\biggr],
\label{H3t}
\end{equation}
where, recalling the definition of $\pi$, from (\ref{2to3})  
we have $\tilde{\pi} = \mu'$.

It is clear from the comparison of these results with the ones reported in Section II and III
 that there is a one-to-one correspondence between the Hamiltonians 
(and the actions) written in the case of tensor modes and those obtained in the 
case of  scalar modes. In particular,  Eqs. (\ref{H1t}) and  (\ref{H2t}) will have their scalar 
counterpart in Eqs. (\ref{hamscal1}) and (\ref{hamscal2}), while  (\ref{H3t}) corresponds to 
(\ref{hamscal3}).

\end{appendix}

\newpage


\end{document}